\documentclass[
    ,final            
  ]
  {aipproc}

\layoutstyle{6x9}

\newcommand{\spitzer}{\emph{Spitzer}}
\newcommand{\msun}{M$_\odot$}

\newcommand{\micron}{$\mu$m}
\newcommand{\apj}{ApJ}
\newcommand{\apjs}{ApJS}
\newcommand{\aj}{AJ}
\newcommand{\aap}{A\&A}
\newcommand{\aaps}{A\&AS}
\newcommand{\mnras}{MNRAS}
\newcommand{\pasp}{PASP}

\begin{document}

\title{Spitzer/IRAC Observations of AGB stars}


\author{Massimo Marengo}{
  address={Harvard-Smithsonian Center for Astrophysics, Cambridge, MA}
}

\author{Megan Reiter}{
  address={University of California, Berkeley, CA}
}

\author{Giovanni G. Fazio}{
  address={Harvard-Smithsonian Center for Astrophysics, Cambridge, MA}
}

\begin{abstract}
  We present here the first observation of galactic AGB stars with
  the InfraRed Array Camera (IRAC) onboard the
  \spitzer{} Space Telescope. Our sample consists of 48 AGB stars of
  different chemical signature, mass loss rate and variability
  class. For each star we have measured IRAC photometry and
  colors. Preliminary results shows that IRAC colors are sensitive to
  spectroscopic features associated to molecules and dust in the AGB
  wind. Period is only loosely correlated to the brightness of the
  stars in the IRAC bands. We do find, however, a tight period-color
  relation for sources classified as ``semiregular'' variables. This
  may be interpreted as the lack of warm dust in the wind of the
  sources in this class, as opposed to Mira variables that show higher
  infrared excess in all IRAC bands.
\end{abstract}

\maketitle


\section{Introduction}\label{sec:intro}

The \spitzer{} Space Telescope \cite{werner2004}, and in particular
its InfraRed Array Camera (IRAC, \cite{fazio2004}), is an ideal
facility to study the distribution of AGB stars in our own and other
galaxies, due to its proficiency in surveying vast areas of the sky in
a short time and its ability to detect sources with infrared
excess. The IRAC colors of AGB stars, however, are not known very
well. The Infrared Space Observatory (ISO) Short Wavelength
Spectrometer (SWS, \cite{valentjin1996}) observed a sample of AGB
stars, finding numerous molecular absorption features in the spectral
region covered by the IRAC photometric system \cite{waters1998}. Just
to mention a few, H$_2$O, SiO, CO$_2$, CO and silicate dust features
have been detected in stars with atmospheric C/O ratio $<1$, while
C$_2$H$_2$, HCN, CS, C$_3$ and carbonaceous dust (SiC and amorphous
carbon) have been found in carbon stars, having C/O $>1$. The presence
and strength of these spectral features depends on the chemistry of
the stellar atmosphere \cite{sloan1998a,sloan1998b} and the mass loss
rate, and can change with time \cite{onaka2002} because AGB stars are
Long Period Variables generically classified as Mira, semiregular
or irregular pulsators.

As a result of dust and molecular features, the IRAC colors of AGB
stars can be quite different from the ``reddened photospheres'' that
one would expect for mass-losing giants of late spectral
type. Simulations we made, based on the available ISO SWS spectra
\cite{marengo2006}, show that AGB stars with different chemical
signature can in some case be differentiated on the basis of the IRAC
colors alone. IRAC colors also appear to be sensitive to the mass loss
rate of the sources, which has also been noted by
\cite{groenewegen2006}. Given the importance of reliably identifying
AGB stars in already available IRAC data from galactic and
extragalactic surveys, the need to study in detail the color
variations of AGB stars is clear.

With this in mind we proposed the observations of a representative
sample of well characterized AGB stars as part of the Cycle-3 IRAC
Guaranteed Time Observations. The goal of the program was to obtain
high-precision IRAC absolute magnitudes and colors. The sample was
chosen to explore the variations of IRAC photometry with time,
variability class, chemical type and mass loss rate. This will enable
longer term goals such as identifying AGB stars in other IRAC
datasets, allowing more detailed studies of the AGB phase as part of
galactic population synthesis models and studies of the galactic
chemical evolution of the diffuse matter in the interstellar
medium. Reliable prescriptions for discriminating AGB stars from other
classes of red sources will also help other projects whose samples are
contaminated by background AGB sources. Given that a fraction of our
targets ($\sim 60$\%) have been observed with ISO, and their SWS
spectra are publicly available, our observations will also provide an
independent dataset for the cross calibration between IRAC and the ISO
SWS. This is particularly important in the case of ``red'' point
sources as mass losing AGB stars, since IRAC has been absolutely
calibrated by using a sample of ``bluer'' stars of spectral type A
\cite{reach2005}. The results of this program will be used to improve
the absolute calibration of IRAC for red stellar sources.

The program will be completed in the first months of 2008, and we are
presenting here the first results derived from $\sim 74$\% of the
total scheduled observations. Once the whole sample of stars will be
observed, the catalog will be made available to the community, with
the goal of aiding the identification of AGB stars in already
available and future IRAC survey.

\section{Target Selection}\label{sec:targets}

Our target list was selected from a number of AGB star catalogs
available in the literature (\cite{guandalini2006, heras2005,
  adelman1998, loup1993, kerschbaum1996} and references therein) with
the intention of representing all main categories of AGB stars. An
important requirement for the selection was the availability of 
reliable distance estimates, to allow for the measurement of absolute
magnitudes. The distances for our target stars were derived either
from astrometric \cite{perriman1997} or interferometric
\cite{millan-gabet2005, vanbelle2002} observations or from models of
radio emission and bolometric luminosity \cite{loup1993}. The distance
of our selected targets varies from 0.17 to 4.1~kpc.

The sample is divided in 4 categories: 22 O-rich AGB stars (MIII
spectral type), 7 intrinsic S stars, 19 carbon stars and 4 M-type cool
supergiants. The supergiants have been added as a comparison sample of
red mass-losing stars with O-rich chemistry outside the AGB. In each
category we have a similar number of Mira, semiregular and irregular
variables, with period ranging from 50 to 822 days. The mass loss
rates of our target stars (estimated from radiative transfer modeling
of their circumstellar emission, see \cite{guandalini2006} and
references therein) range from $10^{-8}$ to $10^{-4}$~\msun/yr in each
category.

Given the variability of the sources, we have requested two epochs
(six months apart, as constrained by the \spitzer{} visibility
windows) for each target, in order to check for variations in the IRAC
photometry on timescales of several months. As of November 2007, 48
sources have been observed in at least one epoch (with 4 carbon stars
missing), with 29 targets (56\% of the total) observed in both epochs.

\section{Data Processing}\label{sec:proc}

The processing of this dataset presented peculiar challenges, due to
the brightness of the sources. Even though we selected our targets to
have a K magnitude larger than 3, to avoid excessive saturation of the
IRAC images, all our stars still saturate the IRAC arrays. The choice
of AGB stars in this brightness range was motivated by the necessity
of having nearby targets, the ones for which a reliable estimate of
the distance is available. Each source was observed with 5 dithered
2~sec full frame exposures, to limit excessive saturation while
avoiding sub-array exposure that, while in many cases still saturated,
have less field of view for PSF fitting photometry.

Starting from the Basically Calibrated Data (BCD) provided by the
\spitzer{} Science Center, we have produced a mosaic image for each
source, in each band and for each one of the two epochs, separately,
using our own post-BCD software \emph{IRACproc}
\cite{schuster2006}.

Due to saturation, aperture photometry was not feasible, so we used a
Point Spread Function (PSF) fitting technique developed specifically
to treat IRAC images with heavy saturation. As described in
\cite{marengo2006}, we have generated a high dynamic range image of
the IRAC PSF in each band, by combining the individual images of a set
of stars with different brightness (Sirius, Vega, Fomalhaut,
$\epsilon$~Eridani, $\epsilon$~Indi and the IRAC calibrator
BD$+$681022). This PSF is rescaled to the actual image of Vega,
providing an absolute photometric reference. By fitting the
unsaturated parts (diffraction spikes and PSF ``tails'') of each
target star image, we have measured the flux ratio between the stars
in our sample and Vega, which we have then converted into Vega
magnitudes for each source. The typical uncertainty of this procedure
is of the order of 1--3\%, better than the typical absolute aperture 
photometry of unsaturated stars with IRAC.

\begin{figure}
  \includegraphics[width=0.55\textwidth,angle=-90]{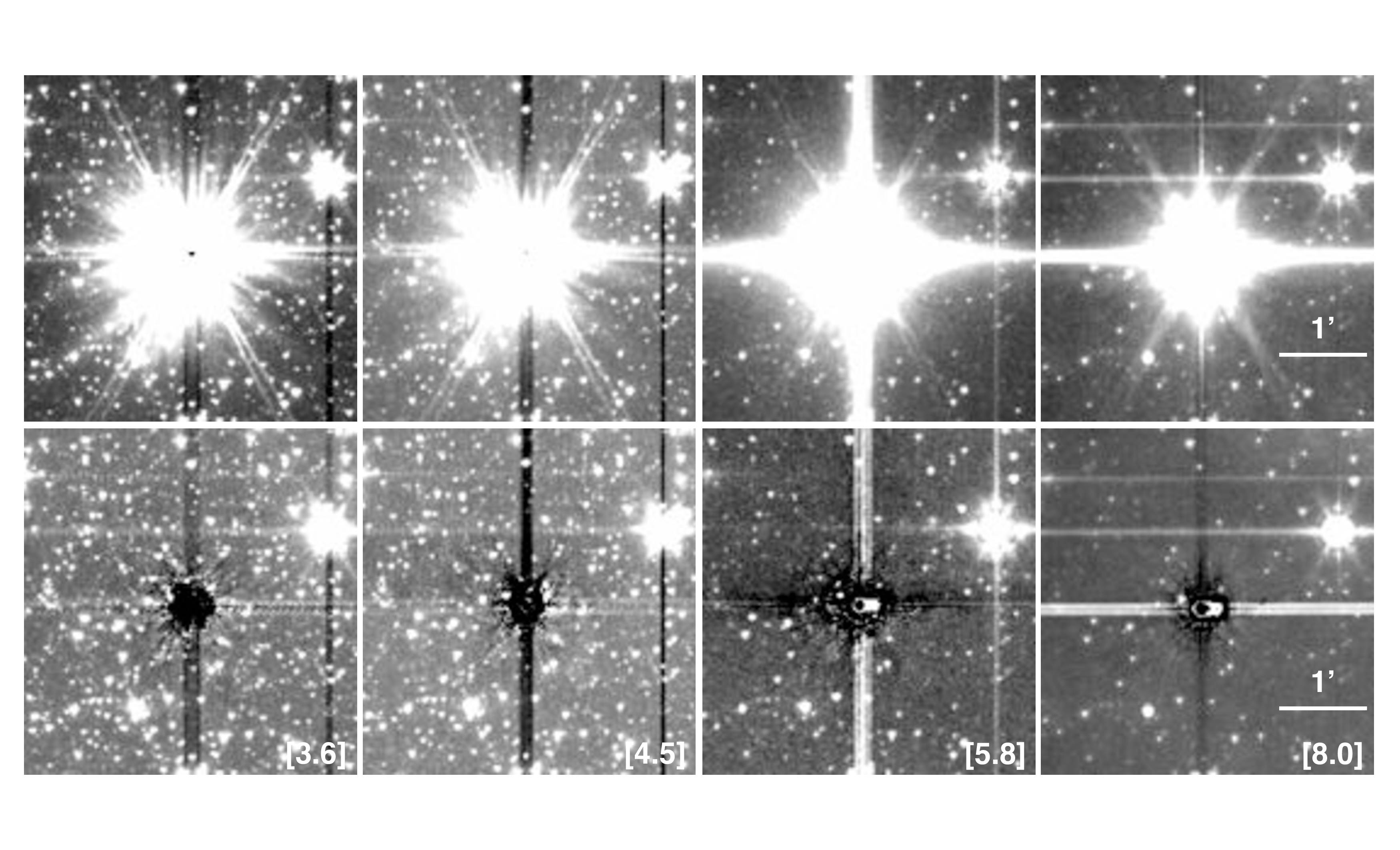}
  \caption{Images of the M-type AGB star CZ~Ser in each of the IRAC
    bands before (top panels) and after (bottom panels) PSF
    subtraction.}\label{fig:1}
\end{figure}

\begin{figure}
  \includegraphics[width=0.60\textwidth,angle=-90]{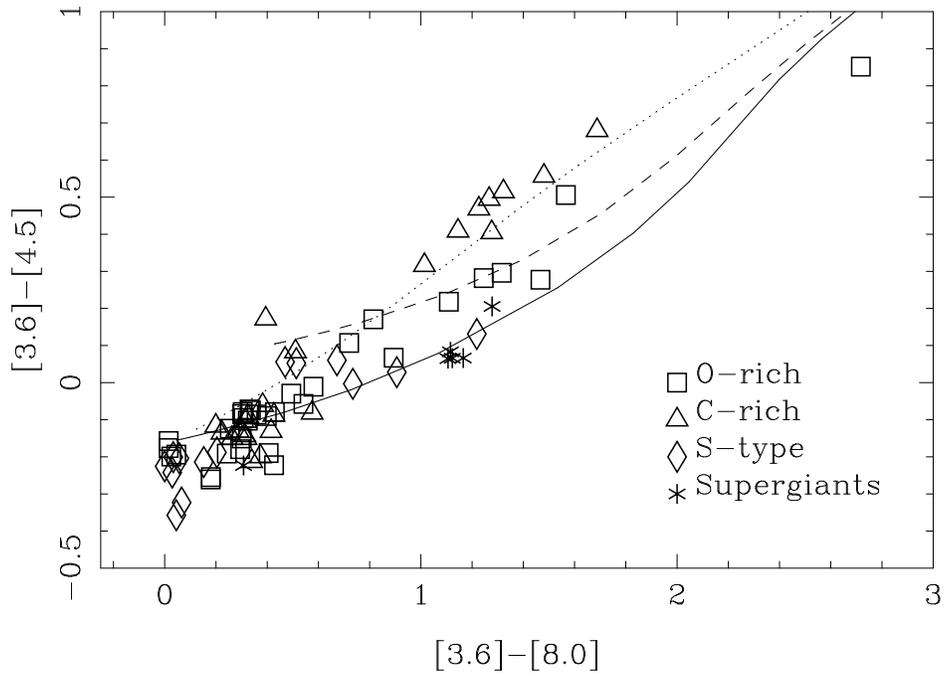}
  \caption{IRAC [3.6]-[8.0] vs. [3.6]-[4.5] color diagram of all the
    available observations to-date. The tracks are models from
    \cite{groenewegen2006}. Dotted line is for carbon stars with
    carbonaceous circumstellar dust of increasing optical depth. Solid
  and dashed line are for silicate envelopes around stars of M0 III and
  M10 III spectral type respectively.}\label{fig:2}
\end{figure}

\begin{figure}
  \includegraphics[width=0.60\textwidth,angle=-90]{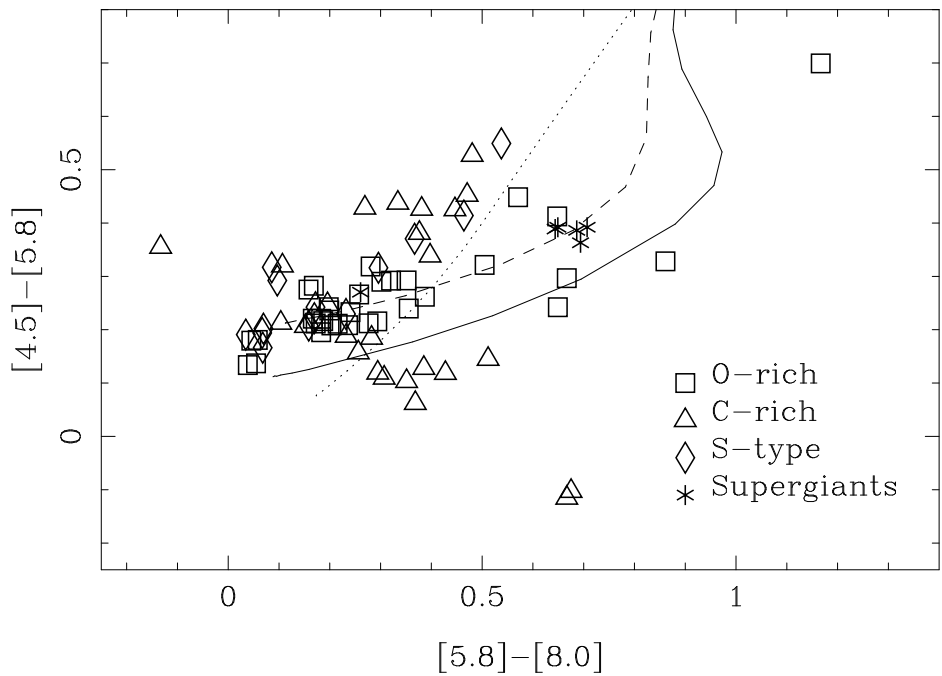}
  \caption{Same as Figure~\ref{fig:2}, but with colors including the
    IRAC 5.8~\micron{} band, causing dispersion of ``naked
    photospheres'' carbon stars from the sequence of carbon stars with
    higher mass loss.}\label{fig:3}
\end{figure}

\section{IRAC Colors of AGB Stars}\label{sec:colors}

Figure~\ref{fig:2} and \ref{fig:3} show IRAC colors for all the
available observations as of November 2007. The sources for which both
epochs are already available have been represented by two separate
points in the diagrams.

As shown in \cite{marengo2006} using synthetic IRAC colors derived
from ISO SWS spectra, and by \cite{groenewegen2006} using radiative
transfer modeling of dusty circumstellar envelopes, the IRAC colors
appear to be sensitive to the optical thickness of the dusty shells
surrounding the sources. The [3.6]-[8.0] color, in particular,
provides the largest spread in function of the optical thickness of
the envelope, due to the thermal radiation from the circumstellar
dust, source of the infrared excess which ``flattens'' the spectrum in
the IRAC wavelength range. The [3.6]-[4.5] color shows a similar
trend, and the combination of these two colors describes a sequence of
increasing excess closely following the model tracks derived by
\cite{groenewegen2006}. The diagrams show a separation of carbon stars
from stars with silicate envelopes (M and S AGB stars, and red
supergiants) for sources with [3.6]-[8.0] color in excess to
1~mag. Inspection of ISO SWS spectra of individual AGB stars with
different chemistry shows that this is caused by two simultaneous
factors: sources with higher excess present a growing 10~\micron{}
silicate feature that is partially contributing to the 8.0~\micron{}
IRAC band, while at the same time the 3.6~\micron{} flux is depressed
by deeper H$_2$O features. Carbon stars, lacking any prominent dust
feature in the IRAC wavelength range, for the same amount of
[3.6]-[4.5] color have a smaller [3.6]-[8.0] excess, causing them to
be aligned on a sequence above the sources with O-rich chemistry. The
infrared excess, for sources with [3.6]-[8.0] $>$ 1, correlates well
with the mass loss rate of the individual stars, albeit with a large
scatter which can be attributed to the large uncertainties in the mass
loss rate estimates.

Figure~\ref{fig:3} shows a trend similar to Figure~\ref{fig:2}, with a
twist: carbon stars with very small mass loss rate align on a separate
sequence with bluer [4.5]-[5.8] and redder [5.8]-[8.0] colors. This was
already shown in \cite{marengo2006}, and attributed to a broad
absorption feature depressing the 5.8~\micron{} flux. Based on
\cite{waters1998} we give a tentative identification of the feature as
being C$_3$ developing in the atmospheres of these dust poor carbon
stars. The feature disappears in stars with thicker circumstellar
envelopes, either because it is filled by the continuum dust emission,
or because the molecule responsible is depleted. As noted in
\cite{marengo2006}, this feature appears to be transient, as some
sources observed with ISO in multiple epochs do not show it in all
spectra. This may be an indication that the feature variability can be
related to changes in the abundances of the molecule in the stellar
atmosphere as the star pulsate, rather than changes in the dust
content of the circumstellar envelope (which is unlikely to show large
scale variations on the short time-scales of the ISO repeated
observations).

\begin{figure}
  \includegraphics[width=0.60\textwidth,angle=-90]{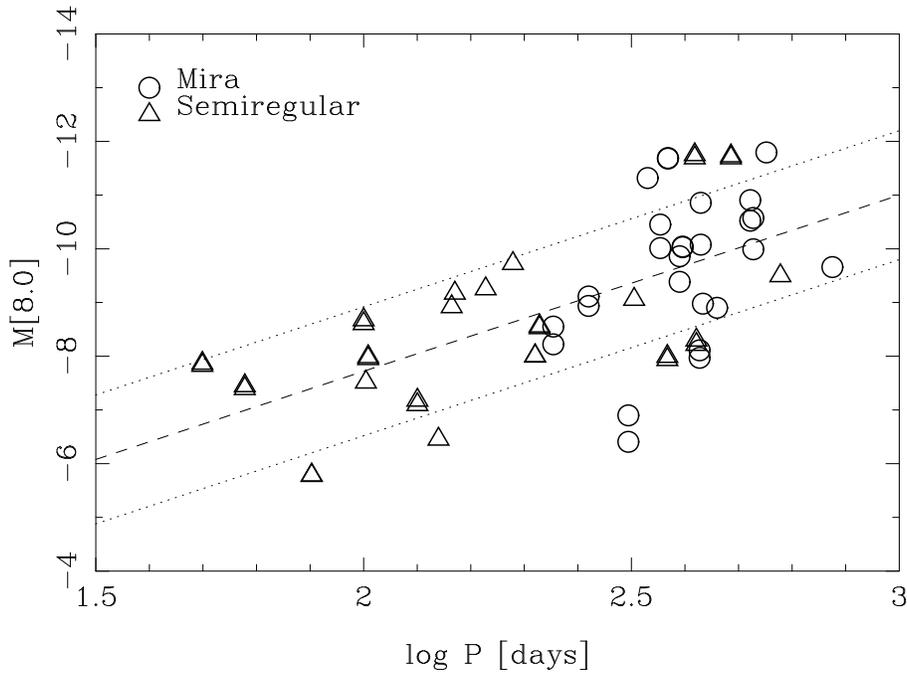}
  \caption{Period - 8.0~\micron{} absolute magnitude relation for all
    the observations available to-date. Circles are sources classified
    as Mira variables in the GCVS \cite{kholopov1988}. Triangles are
    sources classified as semiregular. The solid line is the best
    linear fit for all data, with $RMS$ dispersion of $\sim 1.2$~mag
    indicated by the two dotted lines}\label{fig:4}
\end{figure}

\begin{figure}
  \includegraphics[width=0.60\textwidth,angle=-90]{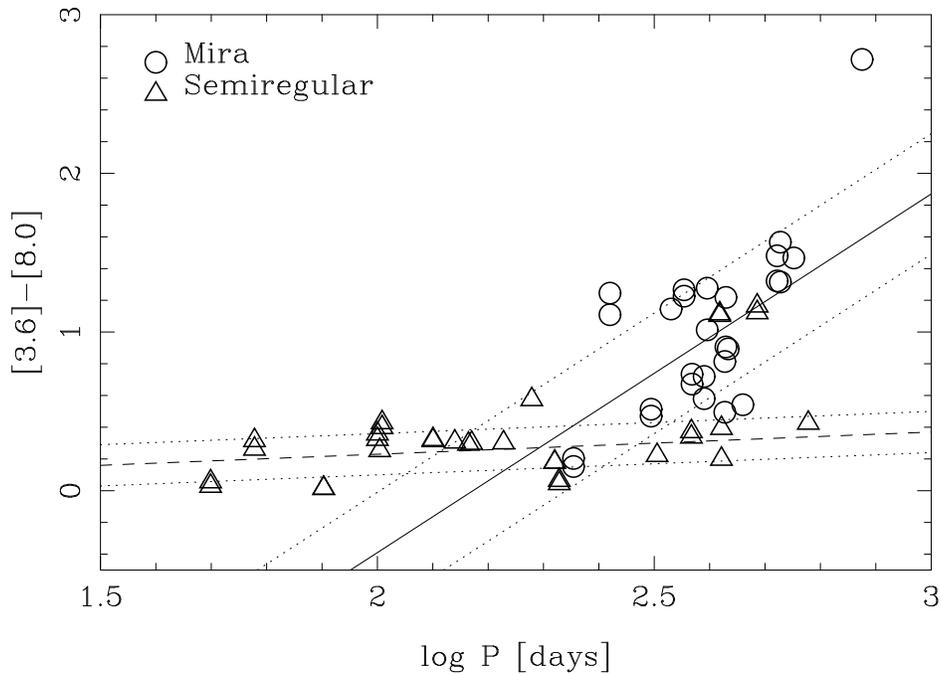}
  \caption{Period - [3.6]-[8.0] color relation for all the
    observations available to-date. Symbols are as in
    Figure~\ref{fig:4}. The dashed line is the best fit for the
    color - period relation of semiregular stars, while the best fit
    for Miras is indicated by a solid line. The dotted lines show the
    dispersion in the best fit relations.}\label{fig:5} 
\end{figure}

\section{IRAC Colors and Variability}\label{sec:period}

Figure~\ref{fig:4} shows the 8.0~\micron{} absolute magnitude of all
sources plotted against their Mira (circles) or semiregular
(triangles) period. Again, sources already observed in both epochs
have separate points, one above the other (same period, different
brightness). Note that all semiregular variables (characterized by
small amplitude) show very little change between the two IRAC
epochs. Only Miras (larger amplitude and longer periods) show
variations that can be as high as $\sim 1$~mag. The plot shows that
there is trend between the period and the luminosity at 8.0~\micron{}
(dashed line), with longer period sources (generally Mira variables)
having higher brightness. The dispersion of $\sim 1.2$~mag may be a
consequence of the varying amount of mass loss of the different stars,
not necessarily correlated to their period, responsible for different
degrees of infrared excess. Diagrams made with other IRAC bands show
a similar trend.

The IRAC period - color relation of all the observed sources is
instead shown in Figure~\ref{fig:5}. As color we have chosen the
[3.6]-[8.0] combination, which is the most sensitive to dust thermal
emission from the circumstellar envelope. Again, Mira variables are
plotted with a different symbol than semiregulars. We have derived
separate linear fits for the period - color relation of the two
classes. Mira and semiregulars show a very different trend: the former
have a very steep relation (solid line) with a large dispersion ($\sim
0.38$~mag), while the latter follow a very tight ($RMS \sim 0.13$~mag)
trend (dashed line).

\section{Discussion and Conclusions}\label{sec:concl}

The IRAC photometry of the observed sources is in general agreement
with models of AGB photospheres surrounded by dusty circumstellar
envelopes of different optical depths, as modeled by
\cite{groenewegen2006}. As expected, the main trend shown by IRAC
colors is an increasing infrared excess for larger optical thickness
of the dust shell, and separation between sources with silicate or
carbonaceous dust. With enough photometric accuracy, however, the data
show deviations from the models that can be attributed to molecular
spectral features in the stellar atmosphere or wind. These features
show time variability when observed at different phases of the star.

The IRAC absolute magnitudes and colors are also correlated with the
Mira or semiregular period of the stars. In general longer period
variables are brighter, but there is a rather large spread in this
relation, probably due to the variations in mass loss rate, and thus
infrared excess, generated by the dusty envelope.

While the overall mass loss rate distribution of Mira and semiregular
variables in our sample is similar, the period - color relation shows
different trends. This may be interpreted as lack of hot dust around
semiregular variables which translates in a smaller infrared
excess. In \cite{marengo2001} a similar result was suggested by the
shape of the silicate feature in a large sample of O-rich AGB stars,
and was interpreted as a sign of time-variability in the silicate dust
production in semiregulars. The new IRAC data supports this
interpretation and extends it to carbon stars (making more unlikely
that this effect is due to differences in the dust composition between
semiregulars and Miras).

The GCVS \cite{kholopov1988} Mira/semiregular classification is
somewhat outdated, and a more accurate description of the AGB
variability should be done in terms of pulsational modes and
amplitudes. In these terms we can rephrase the above conclusions by
noticing that semiregular variables are the ones with smaller
amplitudes, while Miras have in general longer periods and
amplitudes. This may imply that the stars with larger amplitudes may
have more efficient dust production rates, resulting in larger
quantities of hot dust in proximity of the stellar photosphere, which
are responsible for the observed larger IRAC infrared excess.

The observing program is scheduled for completion for the first months
of 2008. Once the data for all stars, both epochs, will be available,
the complete dataset will be published as template photometry of
galactic AGB stars, for general use of the community.




\begin{theacknowledgments}
This work is based on observations made with the \spitzer{} Space
telescope, which is operated by the Jet Propulsion Laboratory,
California Institute of Technology, under a contract with
NASA. Support for this work was provided by NASA through an award
issued by JPL Caltech. This work is also supported by the National
Science Foundation's Research Experience for Undergraduates program.
\end{theacknowledgments}



\bibliographystyle{aipproc}   


\begin{thebibliography}{29}
\bibitem{werner2004} Werner, M.W. et al. 2004, \apjs{} 154, 1
\bibitem{fazio2004} Fazio, G.G. et al. 2004, \apjs{} 154, 10
\bibitem{valentjin1996} Valentjin et al. 1996, \aap{} 315, 60
\bibitem{waters1998} Waters, L.B.F.M. et al. 1998, in proc. ``The
  Universe as seen by ISO'', (ESA SP-427), p. 219
\bibitem{sloan1998a} Sloan, G.C. \& Price, S.D. 1998, \apjs{} 119, 141
\bibitem{sloan1998b} Sloan, G.C., Little-Marenin, I.R. \& Price,
  S.D. 1998, \aj{} 115, 809
\bibitem{onaka2002} Onaka, T., de Jong, T. \& Yamamura, I. 2002,
  \aap{} 388, 573
\bibitem{marengo2006} Marengo, M. et al. 2006, in proc. ``Why Galaxies
  Care about AGB Stars'', Vienna 7-11 August 2006, arXiv:astro-ph/0611346
\bibitem{groenewegen2006} Groenewegen, M.A.T. 2006, \aap{} 448, 181
\bibitem{reach2005} Reach, W.T. 2005, \pasp{} 117, 978
\bibitem{guandalini2006} Guandalini, R. et al. 2006, \aap{} 445, 1069
\bibitem{heras2005} Heras, A.M. \& Hony, S. 2005, \aap{} 439, 171
\bibitem{adelman1998} Adelman, S.J. \& Maher, D.W. 1998, IBVS 4591, 1
\bibitem{loup1993} Loup et al. 1993, \aaps{} 99, 291
\bibitem{kerschbaum1996} Kerschbaum, F. \& Hron, J. 1996, \aap{} 308,
  486
\bibitem{perriman1997} Perriman, M.A.C. et al. 1997, \aap{} 323, L49
\bibitem{millan-gabet2005} Millan-Gabet et al. 2005, \apj{} 620, 961
\bibitem{vanbelle2002} van Belle, G.T., Thompson R.R. \&
  Creech-Eakman, M.J. 2002, \aj{} 124, 1706
\bibitem{schuster2006} Schuster, M., Marengo, M., Patten, B. 2006,
  SPIE Meeting, Orlando, \#6720-65
\bibitem{marengo2006'} Marengo, M., Megeath, S.T., Fazio, G.G.,
  Stapelfeldt, K.R., Werner, M.W. \& Backman, D.E. 2006, \apj{} 647,
  1437 
\bibitem{kholopov1988} Kholopov, P.N. et al. 1988, \emph{General
    Catalog of Variable Stars}, $4^{th}$ edn. 1985 Nauka, Moscow
\bibitem{marengo2001}  Marengo, M., Ivezic, Z. \& Knapp G.R. 2001,
  \mnras{} 324, 1117 
\end{thebibliography}



\end{document}